# EXIST: mission design concept and technology program

J. E. Grindlay[*#a], W. W. Craig[b], N. Gehrels[c], F. A. Harrison[d], J. Hong[a]

[a]Harvard College Observatory, Harvard-Smithsonian CfA, Cambridge, MA
[b]Lawrence Livermore National Lab, Livermore, CA
[c]NASA Goddard Space Flight Center, Greenbelt, MD
[d]California Institute of Technology, Pasadena, CA

## ABSTRACT

The Energetic X-ray Imaging Survey Telescope (EXIST) is a proposed very large area coded aperture telescope array, incorporating 8m$^2$ of pixellated Cd-Zn-Te (CZT) detectors, to conduct a full-sky imaging and temporal hard x-ray (10-600 keV) survey each 95min orbit. With a sensitivity (5$\sigma$, 1yr) of ~0.05mCrab (10-150 keV), it will extend the ROSAT soft x-ray (0.5-2.5keV) and proposed ROSITA medium x-ray (2-10 keV) surveys into the hard x-ray band and enable identification and study of sources ~10-20X fainter than with the ~15-100keV survey planned for the upcoming *Swift* mission. At ~100-600 keV, the ~1mCrab sensitivity is 300X that achieved in the only previous (HEAO-A4, non-imaging) all-sky survey. EXIST will address a broad range of key science objectives: from obscured AGN and surveys for black holes on all scales, which constrain the accretion history of the universe, to the highest sensitivity and resolution studies of gamma-ray bursts it will conduct as the Next Generation Gamma-Ray Burst mission. We summarize the science objectives and mission drivers, and the results of a mission design study for implementation as a free flyer mission, with Delta IV launch. Key issues affecting the telescope and detector design are discussed, and a summary of some of the current design concepts being studied in support of EXIST is presented for the wide-field but high resolution coded aperture imaging and very large area array of imaging CZT detectors. Overall mission design is summarized, and technology development needs and a development program are outlined which would enable the launch of EXIST by the end of the decade, as recommended by the NAS/NRC Decadal Survey.

## 1. INTRODUCTION

The sky has not been surveyed at high sensitivity, or high spatial-spectral-temporal resolution, to fully explore the extreme thermal or non-thermal universe which meet in the hard x-ray/soft $\gamma$-ray band, ~10-600 keV. Indeed, the only full-sky survey to date was performed by the non-imaging HEAO-A4 experiment in 1979 which achieved a 13-180 keV flux sensitivity limit of about 0.05 – 0.3X that of the brightest cosmic hard x-ray source (the Crab Nebula) and modest resolution. The resulting hard x-ray source catalog reported by Levine et al[1] contained some 80 sources, essentially all of which were previously known from more sensitive medium x-ray (2-10 keV) surveys. In contrast, the (true) imaging soft x-ray (0.5-2.5keV) survey carried out in 1991 with the ROSAT grazing incidence telescope catalogued[2] some 80,000 sources full sky with a flux (in band) sensitivity of ~0.05mCrab (or ~1 x 10$^{-12}$erg/cm$^2$-sec) vs. the ~50 mCrab value reached in the 13-80 keV band by HEAO-A4. The "mCrab" flux unit is a convenient observatory-independent measure of detector sensitivity in units of 10$^{-3}$ of the Crab nebula flux and may be converted to an energy flux $F_x$ for a band of multiplicative width W = $E_{mzx}/E_{min}$ (>1) by approximating the Crab x-ray photon spectrum as a dN/dE = 10 E$^{-2}$ photons/ cm$^2$-sec-keV, which then gives $F_x = F_{mCrab} [\ln(W) \cdot (1.6 \times 10^{-11})]$ erg/cm$^2$-sec. Thus 0.05 mCrab corresponds to 5 x 10$^{-13}$ erg/cm$^2$-sec for W=2, as achieved for the 5-10

---

keV band in the HELLAS survey² for active galactic nuclei (AGN) conducted with the MECS instrument on BeppoSAX.

In this paper we describe the "ultimate" all-sky survey possible for the hard x-ray band, that proposed for the Energetic X-ray Imaging Survey Telescope (EXIST). The EXIST mission was originally proposed in the 1994 NASA call for New Mission Concepts and selected (along with GLAST and what became Constellation-X) for study. It was recommended by both the 1999 Gamma Ray Astrophysics Program Working Group (GRAPWG) and, particularly, by the NAS/NRC 2001 Decadal Survey for Astronomy and Astrophysics (AASC), which recommended it be launched by 2010 to carry out a very high sensitivity all sky survey as well as be the Next Generation Gamma Ray Burst mission. Here we describe the current mission concept, as established by two engineering design reviews at NASA/GSFC in May 2000 and November 2001. EXIST is now conceived as a very large area ($8m^2$) array of imaging solid-state detectors (CdZnTe; CZT) which read out 3 wide field (60° x 75°) and adjacent coded aperture telescopes, with combined field of view (FoV) of 180° x 75°. With zenith pointing, the scanning fan-beam Observatory (cf. Figure 1) will image the full sky each 95min orbit and achieve a flux sensitivity of ~0.05mCrab (5σ, ~1y). It will be launched by a Delta IV as a freeflyer (intermediate class) mission. The mission concept studied originally for implementation on the ISS[4] would have reduced observing time due to higher backgrounds, smaller total FoV (due to ISS structures), and with its fixed mounting could not carry out simultaneous pointing and survey observations.

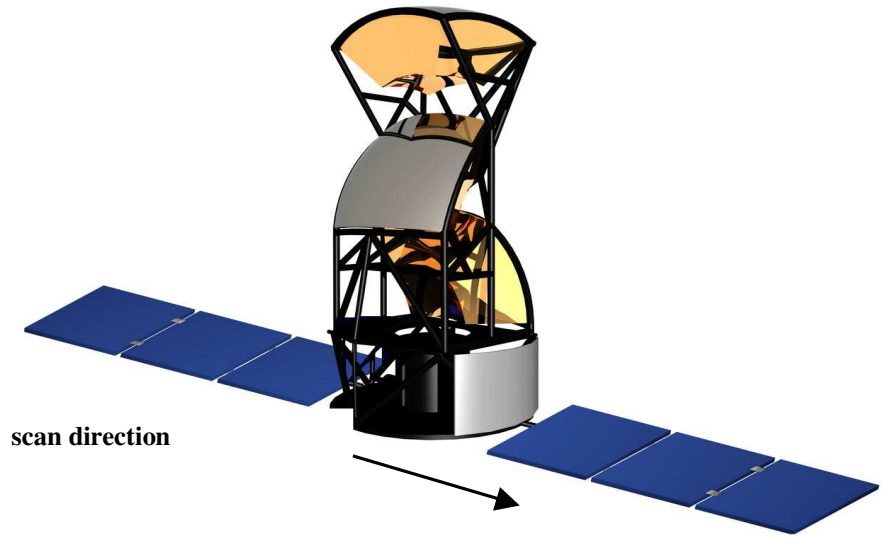

**scan direction**

Figure 1: EXIST overview, showing three coded aperture telescopes stacked on spacecraft maintaining Zenith pointing to scan the full sky each orbit. Overall envelope for telescope and S/C is 9.6m x 4.8m.

We begin with a brief overview of the science goals and drivers for the mission and then summarize the most recent mission study results. We conclude with a discussion of the essential technology development needed.

## 2. SCIENCE GOALS AND MISSION DRIVERS

**2.1 Science goals: survey and study black holes on all scales**

Although the hard x-ray band is key for the study of a wide range of astrophysical phenomena, black holes are particularly prominent. Their accretion-produced output spectra peak (typically) in the hard x-ray band, with maximum energy flux $F_x \propto \upsilon F_\upsilon$ at photon energies $E_x \sim 100$ keV for most black holes (BHs) on mass scales from the stellar mass BHs in x-ray binaries in the Galaxy to the supermassive BHs which power active galactic nuclei (AGN). Black holes are currently favored as the objects responsible for gamma-ray bursts (GRBs), with the prompt gamma-burst radiation produced by release of a gamma-dominated relativistic jet in the formation of a stellar-mass BH in the collapse of a massive star in a "hypernova" event[5]. GRBs also have their peak luminosity at $E_x$ ~200-300keV (though probably for different reasons).

Thus the primary science goal for EXIST is to survey BHs on all (mass) scales. The study of black holes is one of the three primary quests for the proposed "Beyond Einstein" program of NASA now being formulated in the next Strategic Plan for NASA Space Science. The science goals for a mission to survey BHs, the "BHProbe", were summarized for the NASA Roadmap Subcommittee as follows:

The Black Hole Finder Probe (BHProbe) will perform the first all-sky imaging survey for black holes (BHs) on all scales: from supermassive BHs obscured in the nuclei of galaxies, to intermediate mass (~100-1000 solar mass) holes likely produced by the very first stars, to stellar mass holes in our Galaxy. A wide-field (~5sr) coded aperture telescope array operating in the hard X-ray band (~10-600 keV), where accreting BHs are particularly luminous, is a promising approach since hard X-rays penetrate the veil. Such a survey [e.g. EXIST, as endorsed by the AASC] would also be the first to achieve all-sky hard x-ray imaging every ~95min orbit, allowing BHs to be surveyed in both time and space.

Recent evidence has suggested that a large fraction of massive black holes in the centers of galaxies are obscured by surrounding gas and dust in the nuclear vicinity. Indeed the three closest supermassive BHs are in the nuclei of obscured and optically dull galaxies. BHProbe would make the first census of such massive BHs in the local universe and distinguish them from "starburst" nuclei (in comparably dusty environments) by the hard X-ray spectra and variability unique to a central BH. Such a census is critical to determine if massive BHs are present in all galaxies and were grown by accretion during the galaxy formation epoch, as suggested by the hard X-ray background radiation and general accretion (X-ray) vs. nuclear (starlight) luminosity density of the universe[6].

BHProbe would enable a wide range of fundamental studies of black holes and the extremes of astrophysics:

(1) BHProbe will measure the supermassive BH content of galaxies in the local universe for a wide range of both obscuration and accretion rate. BHProbe can identify the most luminous and thus massive obscured BHs at larger redshifts to constrain the growth rate of massive BHs. NGST could measure the star formation rate in the same obscured AGN for comparison. Followup detailed studies with Constellation-X can measure fundamental BH properties (spin, mass) in the most optimal targets.

(2) BHProbe will perform the first continuous variability survey for BHs in the hard X-ray band, where their luminosity per decade of frequency locally peaks. The inherently unpredictable largest variations of these messy eaters, such as when entire stars are gulped, could be measured as hard X-ray flares (duration hours - days) from the preponderance of obscured AGN in the local universe. Comparisons with LISA would allow the space-time vs. accretion disk (or flow) signatures to be disentangled for infall of compact vs. the more numerous non-degenerate stars. The large galactic population of stellar mass BHs in binaries, which appear as X-ray transients, would be mapped and distinguished from those with neutron stars by their hard X-ray spectral and temporal signatures.

(3) BHProbe will conduct a hard X-ray survey ~1000X more sensitive than the only previous full-sky survey (HEAO-A4) and with ~20X more sensitivity than BATSE for Gamma-ray bursts (GRBs). Increases in sensitivity, energy band coverage, temporal and spectral resolution are all factors of ≥5-10 over those projected for the Swift mission. Thus BHProbe would be the *Next Generation GRB mission*, allowing the most sensitive study of the highest redshift GRBs expected from the formation of intermediate mass BHs at the epoch of formation of the very first (massive) stars. Such objects are probable seeds for massive BH formation, may contribute to dark matter in galaxies, and could also be detected locally by BHProbe when they encounter dense cores of giant molecular cloud complexes and accrete as hard X-ray sources.

(4) BHProbe will at the same time survey other extremes of astrophysics: (a) through long duration timing studies of accreting X-ray binaries, it can improve our understanding of physical processes occurring in the extreme environments of high gravity, high magnetic field, and high radiation energy density; (b) through the detection of hard X-ray nuclear decay lines from $^{44}$Ti and other species, it can provide crucial information on supernova and nova rates in the galaxy, and constrain models of cosmic nucleosynthesis; (c) through high sensitivity studies of soft gamma-ray repeaters, it can constrain the formation and evolution of neutron stars with the most extreme magnetic fields in the local universe; and (d) through measures of the hard X-ray spectra of distant active galactic nuclei in conjunction with high energy gamma-ray spectra from GLAST and VERITAS, it constrain the shape of the diffuse infrared radiation background (and thus star formation rate) as a function of cosmic time.

**2.2 Mission Drivers**

As a coded aperture imaging telescope, EXIST measures each source over the full detector area visible to that source and is thus background limited: by the cosmic diffuse flux below ~100-200 keV (depending on FoV) and by detector internal background contributions at higher energies. Thus the minimum detectable flux scales as $F_{xmin} \propto (B/A\ T)^{0.5}$ and both detector area A and exposure time T must be maximized to achieve the desired ~0.05mCrab sensitivity for the survey. In order to survey sources on the broadest range of timescales, as well as maximize exposure to GRBs and transients, the imaging detector should be wide-field and scan the full sky each orbit. This leads, naturally, to the mission requirements for very large detector area and large FoV on a scanning, nominally zenith pointed, telescope which sweeps out the full sky each orbit. A brief discussion of the sensitivity and scanning requirements, with particular relevance for GRBs, is given elsewhere[7].

The science driver to extend the hard x-ray sensitivity up to at least 300-400 keV (~$E_{peak}$ for both GRBs and AGN in their $\upsilon F_\upsilon$ spectral distributions) and with appreciable sensitivity at the electron-positron annihilation energy 511 keV means that the coded aperture mask must be thick (~7mm Tungsten) and that the imaging CZT detector array must also be relatively thick (≥5mm) and also have depth-sensing capability to minimize mask pixel parallax. The 5-10' angular resolution required to avoid source confusion (adopting the usual ~1/40 source per beam criterion) and allow imaging and location of the >30,000 AGN expected at the desired survey sensitivity $Fx(100-200keV) = 5 \times 10^{-13}$erg/cm$^2$-sec imposes constraints on the coded mask and telescope design. For a coded mask - detector distance (focal length) of 1.5m (constrained to minimize overall telescope mass), the required small (~2.5mm) unit pixel size will impose auto-collimation unless the wide FoV is segmented into adjacent smaller ones and the mask holes themselves are radial in the extended and quasi-hemispherical coded mask required for minimum mass and size. The science and instrument requirements are summarized in Table 1.

**Table 1:** Science goals and instrument-mission requirements

| Parameter | Science goal | Requirements & implementation |
|---|---|---|
| Sensitivity (5σ, ~1y) | 0.05mCrab (10-150keV); ~0.5mCrab (150-600 keV) | 4-8m$^2$ total area of CdZnTe imaging (1.2mm pixels) detector, ≥5 mm thk |
| Angular resolution | 5' (resolve sources); 10" (locations) | 2.5mm mask pixels; 5" aspect |
| Instantaneous FoV | full sky/orbit & ≥20% cont. obs./src | 180$^\circ$ x 75$^\circ$ fully-coded hemi-mask |
| Energy band (>20% QE) | 10-600 keV | ~7mm thick W mask; radial holes; depth-sensing readout |
| Energy, time resolution | 1-4 keV (FWHM); 2μsec | low noise ASIC; event telem. req. |

In the following section we outline the mission concept and current version of the telescope and detector design to achieve these requirements.

## 3. MISSION CONCEPT AND PRELIMINARY DESIGN

We next summarize the current mission design concept from the top down: the telescope layout and imaging design, the shields and collimation, the detector packaging and readout systems, and the data systems and on-board processing.

### 3.1 Telescope and coded mask configuration

The required wide FoV and broad-band energy coverage mean that EXIST should be a scanning, rather than pointed, telescope mission (although, as described below, inertial pointing as an Observatory also achieves most survey and GRB objectives while enabling high sensitivity continuous timing studies of sources in a smaller FoV, and thus lower background, sub-array of the central telescope). A major advantage of scanning is also that the shadowgram unique to each source is continuously averaged over the detector plane to minimize systematic variations in detector response and background variations across the detector that are otherwise dominant sources of systematic noise in a fixed-pointed coded aperture telescope. This leads to a telescope and detector layout as shown in Figures 2 and 3.

The 3 coded aperture telescopes (Left, Right and Upper; cf. Figure 2) are each with total FoV of $60°$ x $75°$ in the **Y** ($\perp$ to ram) x **X** (along ram) directions so that by maintaining local-zenith pointing, the resulting fan beam ($180°$ x $75°$) sweeps out the full sky each orbit. A source on the orbital equator is observed for (nominally) 20% (=75/360), with those towards the orbital poles having increased coverage. The vertical stack of the 3 telescopes is a minimum mass and volume configuration for the desired total detector and coded mask areas ($2.7m^2$ and $7m^2$, respectively) and focal length (1.5m) needed for each telescope. The overall 3-telescope Observatory has dimensions and envelope as shown in Figure 1 and total mission mass of 8500kg, which is dominated by the tungsten coded apertures and shielding. The coded aperture pattern can be constructed as a URA pattern (dimension ~250 x 300), with a 2 x 2 cycle viewed by the $50°$ x $58°$ FoV collimation[§] for each sub-telescope, or be random. Mask segments could be fabricated as either flat or curved W plates (~60cm square), with 2.5mm unit pixel size, by laser cutting the pixel pattern into 1 mm thick W sheets. By slightly offseting the pixels across the seven sheets laminated to fabricate a mask segment, radial alignment of the pixels with the overall mask radius of curvature can be achieved in order to minimize auto-collimation by the mask. The coded mask can be self-supporting by leaving a thin (0.1mm) support grid between adjacent open pixels with minimal (15%) loss of throughput at low energies but negligible loss of sensitivity; or isolated closed mask pixels can themselves (just) be supported. Detailed simulations of the imaging response of this "hemi-spherical" coded aperture imaging system are needed as part of the mission design study, with preliminary results (by J. Hong) indicating near-optimum response.

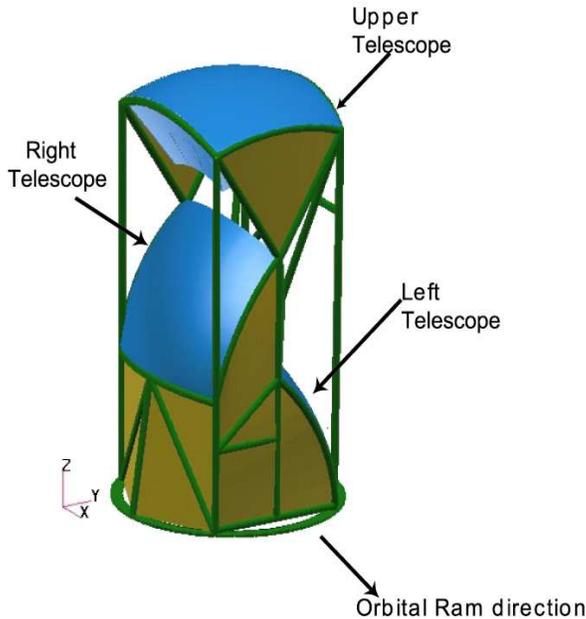 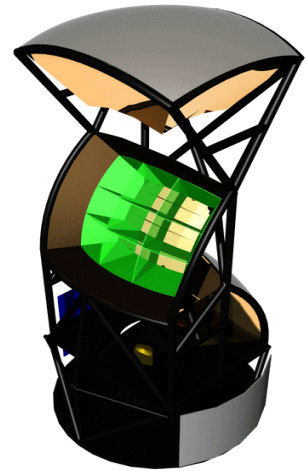

Figure 2: Structural view of 3 coded aperture telescopes for EXIST. Underlying spacecraft shown in Figure 1.

Figure 3: Cutaway view of 3 x 3 sub-telescope array for Right Telescope with active-passive collimator shields (green) each defining $50°$ x $58°$ FoV.

---

[§] The sub-telescope pointing directions are offset from each other (cf. Fig. 3) by $10°$ and $12.5°$ (along their $50°$ and $58°$ axes) so their combined FoV (of the 3 x 3 array) yields the desired total fully-coded FoV of $60°$ x $75°$ for a Telescope.

Since at the high total rates expected (~3 x $10^4$ cts/s) in such a large area detector-telescope array it is most efficient to bring each photon down tagged with position, energy and time, the telescope pointing can be very coarse (~1°). Pointing requirements are nominally set only by keeping the Earth's limb below the edge of the field of the Right or Left telescope. Telescope or S/C roll is similarly coarsely constrained to be maintained within ~5° of the perpendicular to the orbital ram direction. However instantaneous aspect must be measured to ~5" from an off-axis star tracker on each telescope (of which any one that violates sun angle constraints can be turned of) as well as on-board x-ray aspect from bright x-ray sources always in the field of view of some telescope. This ensures source positional accuracies at the ~10" level (systematics limit imposed both by imaging centroiding ability and mask-detector plane structural rigidity) for sources detected at ≥30σ significance.

**3.2 Shielding, collimation and high energy GRB spectroscopy**

The detector and imaging fields of view are designed to achieve maximum sensitivity and imaging response. Monte Carlo simulations of the expected background and imaging sensitivity have been conducted using MGEANT to arrive at the collimation and shielding configuration design concept. The large FoV desired for scanning sensitivity and sky coverage and exposure duty cycle means that the background is dominated by the cosmic diffuse flux up to ~150 keV and both internal detector background (due to cosmic ray spallation and neutron-capture effects) and shield leakage at higher energies. More detailed studies of the high energy backgrounds are needed, though balloon flight background studies with differing thickness imaging CZT detectors (Jenkins et al[8] and references therein) as well as neutron capture background simulations by Harrison et al[9] indicate that internal backgrounds are reasonably well understood. The imminent launch of INTEGRAL[10] will provide the first long-duration measurement of backgrounds in CdTe (the 2mm thick LEGRI detector plane) with a large area coincidence shield (and detector; the underlying thick CsI detector, ISGRI).

A combination of active and passive shielding is needed for EXIST. The large area (2.7m$^2$) CZT array in each of the 3 Telescopes is shielded from below (i.e. the CR-induced background in S/C structures as well as Earth albedo) by a 2cm thick active shield below each sub-telescope (ST) module (0.3m$^2$). Ideally, this would be BGO, for maximum stopping power; design studies are being conducted to assess the relative performance of a CsI(Na) alone or in combination with an external passive shield. The side collimators (Fig. 3) are currently baselined to be CsI (1cm thick) to allow both collimation and active CR-induced background rejection. The upper portion (~1/3) of the collimator walls can be passive, given the small solid angle they present to the detector plane, though detailed design studies are underway.

The rear shield, and possibly the active collimator, will be pulse-height analyzed for data accumulated in a rolling FIFO buffer (>100sec; commandable) for spectroscopy of GRBs up to ~3-10MeV. This will extend the upper energy limit for GRB spectra, with high time resolution, to more completely overlap with GLAST, for which EXIST will obtain precise (~10" – 1') positions. The onboard GRB trigger and initial imaging analysis system will automatically pipe shield spectroscopy data to the output telemetry for the burst duration (in place of imaging CZT data from STs for which the GRB is not visible). To keep shield rates well below lock-up values, and to allow GRB spectroscopy at the high rates expected, each ST will have its own digital and processing unit, with each flat shield piece (collimator sidewall or rear ST shield) having its own digital ID and PHA value. An event in a given ST can be stored (e.g. during a GRB trigger) with any of its adjacent 5 shield data values (e.g. to re-construct Compton scattered events), though for normal survey mode such events are normally rejected onboard as background induced.

**3.3 CZT detector plane: readout and packaging concepts**

Although the very large area of the CZT imaging detector is technically demanding, a modular approach to the detector readout, packaging and data systems will allow prototype development, integration and testing and in-flight redundancy. Here we describe a possible system, currently under study and prototype development, which employs CZT crystals with 2D anode pixel readout connected to a corresponding 2D input array of a custom-designed ASIC which provides the charge preamplifier, shaping amplifier, and sample and hold stage for each pixel that exceeds a commandable threshold. The science requirements for low noise (for good energy resolution and low energy threshold) and a broad energy band (10-600 keV) both point to a 2D pixel, rather than crossed 1D strip, detector readout. Compton scattering (dominant above 200-300 keV) renders strip detectors ambiguous, whereas both energy and position (imaging) can be corrected (in part) for those events for which multiple interaction sites can be recorded. The $N^2$ rather than 2N channel readout for imaging arrays of dimension N x N pixels can be dealt with by low-power ASICs and a flip-chip mounting system which enables vertical stacking of the CZT onto the ASIC readout so that detector assemblies (and their readout) can be close-tiled in a flat array, over large area, with minimal gaps.

The unit cell of the detector is then a detector crystal assembly (DCA) composed of a 2 x 2 array of CZT crystals each 20 x 20 x 5mm (or thicker; up to 10mm) and each close-tiled (0.1mm gaps) onto a single coupling board which routes contacts from the anode pixels of the CZT (on top) to the ASIC input pads (on bottom). The anode pixels on each crystal are a 16 x 16 array of Pt contacts evaporated onto the CZT crystal through a shadow mask with 1.25mm pitch, 1.0mm pad size and thus 0.125mm inter-pixel gaps. The outer-most pixels (around the crystal edge) are truncated to have contact pad width 0.75mm in the direction toward the edge to leave a 0.25mm gap to guard against crystal edge imperfections and leakage currents. A schematic of the DCA and its mounting in a detector module tray is shown in Figure 4.

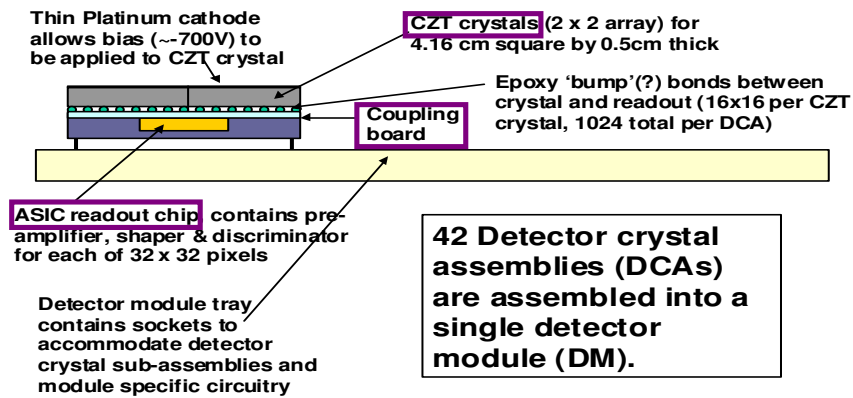

Figure 4. Concept for CZT crystal close-tiling and anode pixel readout by ASIC connected via a coupling board, which provides both mechanical support and pixel de-magnification from CZT (1.25mm pitch) to ASIC (0.6mm pitch).

In order to cover the rectangular FoV (50 ° x 58 °) for each sub-telescope (ST), 6 x 7 such DCAs are combined to form one detector module (DM), and 2 x 2 DMs then comprise a full ST. Digitization of the ASIC output is done at the DM level, along with command and control of each DCA. The DCAs are socket mounted into the DM tray for relatively easy integration and replacement.

In order to optimize imaging with small spatial pixels (1.25mm) over the large FoV and correct events at variable interaction depth for their x,y location on the detector surface, depth detection is needed. This also allows improved energy resolution by allowing corrections for charge collection as a function of interaction depth, which can be derived from a simple cathode/anode pulse height ratio for each event (cf. Narita et al[11] and references therein). A distributed set of separate cathode readout contacts, which also supply the negative detector bias voltage (-700V across a blocking capacitor) to each crystal separately, is possible with an edge-mounted cathode readout board. The board, with on-board ASIC, reads out crystals along either side of an entire row of the 7 DCAs (28 crystals) in a single DM. To allow for the board thickness without occulting detector pixels, as well as to allow for DCA insertion on their mounting board, the DCAs are spaced by one detector pixel width (1.25mm). Six boards, each only ~2cm high, then service an entire DM. A schematic representation is shown in Figure 5 for a prototype 4 x 4 DCA array, rather than the possible 6 x 7 array. The thin PC boards provide additional collimation at low energies (<20keV) in the scan direction to minimize background from very bright x-ray binaries (e.g. Sco X-1 and galactic bulge sources) which would otherwise reduce survey sensitivity in a larger region around the galactic bulge. The board height, and thus low energy collimation, is one of several parameters to be optimized. The shaped-held cathode pulses are digitized off board and combined with their appropriate anode pulse height in the data formatter. An initial prototype version of this system has been tested for a single 20mm crystal on a single DCA (2 x 2 crystals, with 2.5mm anode pixel pitch) and shown to significantly improve energy resolution; results are reported separately by Hong et al[12].

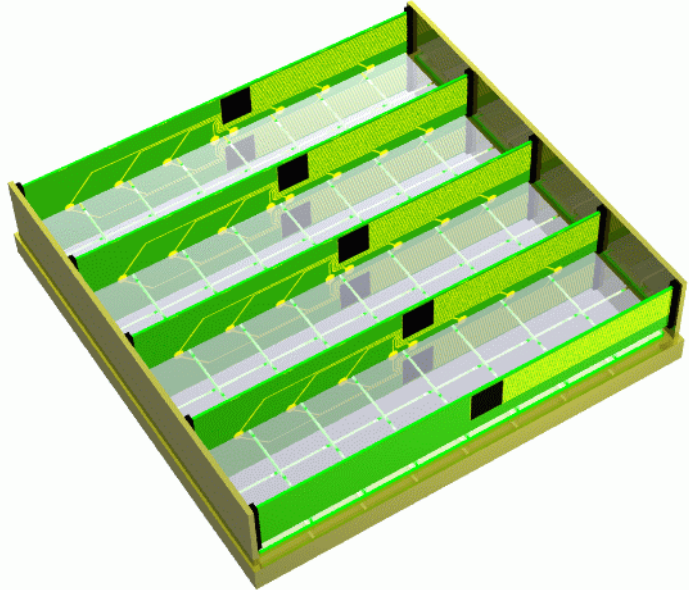

Figure 5. Prototype detector module concept (16cm x 16cm) showing edge-on cathode readout boards for bias (not shown) and cathode signal traces to central ASIC (black rectangle) and output lines to connectors on mounting frame.

**3.4 Data handling systems and on-board processing**

Each detector crystal assembly, 1024 pixels, is read out by an ASIC bonded to the CZT crystals of the DCA through a coupling board. A Detector Module Data Handling Unit (DMDHU) digitizes and collects the data from each of 42 DCAs in a detector module. Each sub-telescope (ST), with 4 DMs, then contains 4 DMDHUs. Each ST is functionally independent and is controlled by a Sub-Telescope Control Unit (STCU). The STCU performs initial processing of the data, receives data from the Shield Interface Unit (SIU), which collects shield triggers (or, upon command, pulse heights) from each of the side shields and rear shield, as described in §3.2, and controls the event veto logic. The STCU also monitors rates for bursts and transients and performs initial positioning of bursts within a sub-telescope field of view. This "de-centralized" control both allows redundancy and also services the wide FoV and separate detector-trigger needs for STs within a single telescope. A given source (e.g. a GRB) is instantaneously viewed by only 4-6 STs so that distributed trigger and data handling is required. The schematic DMDHU and STCU configurations and data paths are shown in Figure 6.

The high data rates expected from the diffuse background detected by the large FoV of each ST (~0.3 cts/cm$^2$-sec, and a factor of ~2-3 larger for bright source regions in the Galaxy), mean that each STCU processes the shield veto for each DM individually: a DM (CZT) rate of ~300-1000 cts/s vs. the total shield rate. To keep the effective rate of the entire 5-sided ST shield, with total area ~$10^4$ cm$^2$ (and thus total trigger rate >10⁷ cts/s) from locking up the DM detector (with ~1μsec time constant for the CZT detector vs. ~0.5 μsec for the shields) then requires the SIU to pre-process shield data. Since most of the high shield rate is due to cosmic ray (CR) events, the total shield rate is first reduced by a factor of ≥2 by single-counting multiple shield coincidence events. Each shield (side panel or rear) has a high vs. low gain discriminator, derived from PMT anode vs. dynode pulses, for tagging photon (<10 MeV) vs. CR (>20-40 MeV) –photon-induced shield events. CZT events coincident with CR events are rejected, whereas those coincident with photon-induced shield events are rejected for onboard GRB and transient analysis but brought down in the full telemetry (as flags) for either rejection or Compton event reconstruction in the full analysis. Remaining (un-tagged) CZT events can also be rejected onboard or in post-processing dependent on CZT energy and interaction depth: e.g. a CZT event at <100 keV with depth in the bottom half of the detector is rejected as likely due to Compton scattering of a background photon incident from below.

Although GEANT simulations of detector backgrounds have been conducted, a more complete set of simulations and veto criteria are needed to develop the trigger and veto systems. A more conservative approach being studied is to reduce the shield segmentation by a factor of 2 in each dimension by having each DM, rather than the 2 x 2 DM array of a ST, be shielded and separately controlled. This offers advantages of complete DM modularity as well as reduced (to 3/4) side-shield mass but the disadvantages of doubling the shield readout system (discriminators, but not PMTs which would be 2 per ST side-shield) and increased dead area (in shields) for the same total detector footprint area.

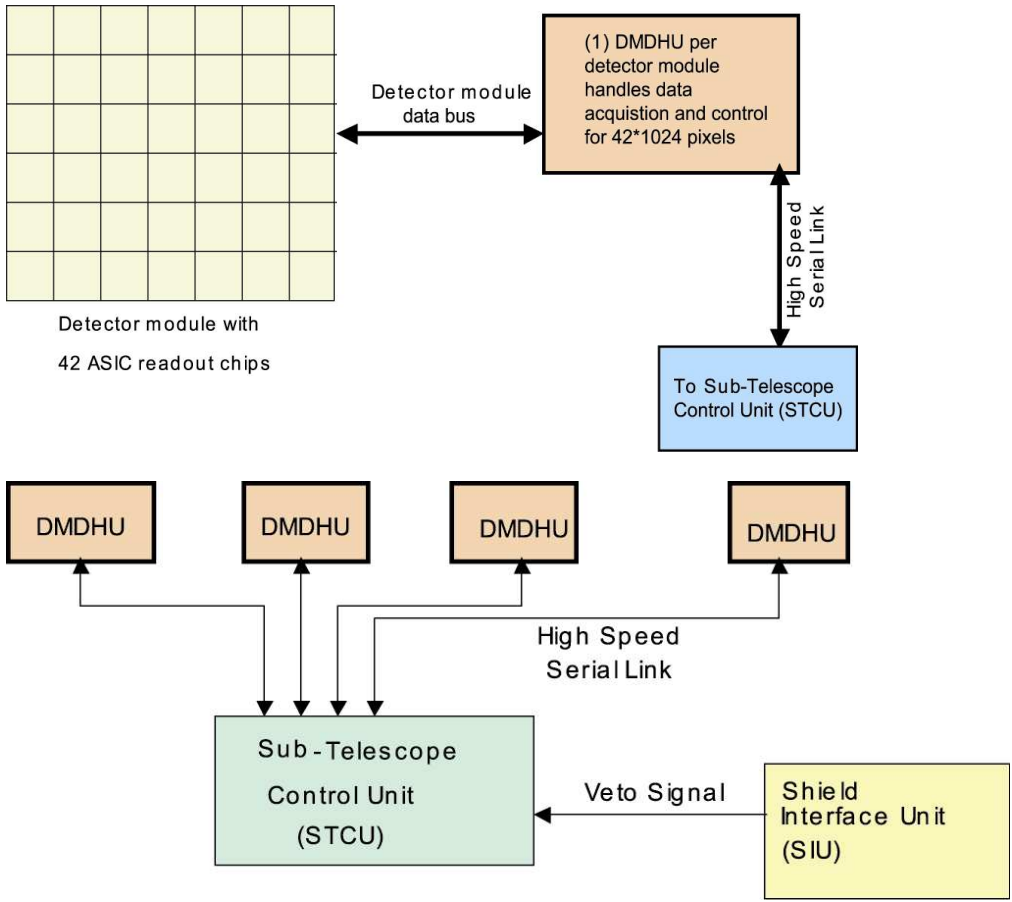

Figure 6. Instrument electronics functional block diagram showing DMDHU (top) and STCU (bottom).

Accepted CZT data, with expected total rate ~3 x 10⁵ cts/s, are formatted in time order so that only 8 bits are needed for ~2μsec absolute timing (vs. the S/C clock, updated by GPS). With event energies (14bits, including 4 bits cathode), positions (22 bits), and shield and status flags (8 bits), the total telemetry needed is 1.5Mbs, which can be brought down via an X-band link every second orbit from on-board digital storage.

On board processing of non-vetoed events is conducted for prompt (<10sec) initial celestial positions and spectra of GRBs. This is done at the Telescope level by comparing instantaneous trigger rates in several broad energy bands between all (9) STs for each Telescope and over several integration timescales. The GRB trigger criteria for initiating prompt image reconstruction processing of buffered data are similar to BATSE although much more flexible. Rates from 4 contiguous STs (which can fully view a given GRB) must exceed threshold(s), and STs must have total rates consistent with a single source scanning across the FoV for the integration time. An initial ~5° celestial position is derived from count rate ratios (again, like BATSE) and for brighter bursts (>10X threshold) refined to ~1° by count rate distributions across the STs. This allows very fast correlation analysis for the actual burst image position since only a small fraction of the large FoV need be analyzed (the full field is processed subsequently). GRB positions to within ~3', which are primarily limited by coarse initial aspect solutions, are then available for transmission via an open low data-rate link through TDRSS to the ground and internet dissemination within ~10sec. With data dumps to the ground nominally every second orbit, final GRB positions (~10" − 1') would be delivered within ~0.5 – 4h, depending on the lapsed time from a GRB to the next data dump.

It may be similarly possible to image longer duration (than most GRBs) transients, such as x-ray novae from accretion disk instabilities in black hole LMXBs, or white dwarf thermonuclear novae (and accompanying 511 keV emission), with onboard processing for (near) real time notification. This might entail correlation images derived from each DM or ST alone, rather than the full Telescope, to allow for faster processing and data transfer with limited flight data processing and storage systems. Given the single-orbit flux sensitivity (see Table 2) of ~2mCrab, it should be possible to do realtime (each orbit) detection and ~1' locations for ≥5 mCrab sources over most of the sky although constraints on system resources have not yet been established.

## 4. SURVEY COVERAGE, SENSITIVITY AND MISSION SUMMARY

Given the mission concept outlined above, we can predict approximate expected backgrounds and exposure coverage and derive predicted sensitivities for continuum flux detection (in bandwidth $\Delta E = E$) or line spectroscopy (with bandwidth $\Delta E = E/50$). The exposure coverage is derived for the full (3-telescope) combined FoV and is shown in Figure 7 for a ~1day interval; survey sensitivities are plotted in Figure 8.

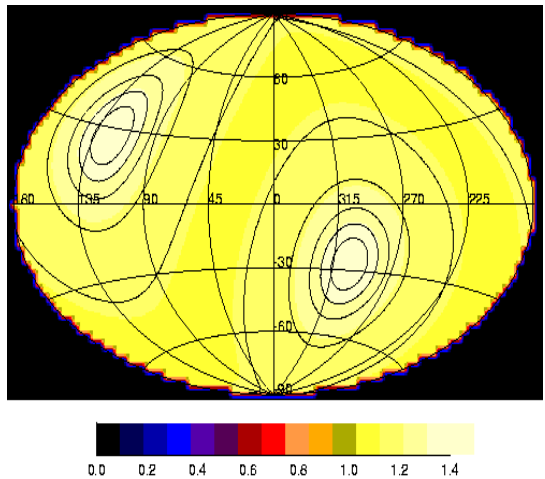

Figure 7. EXIST 1day relative exposure (galactic coordinates).

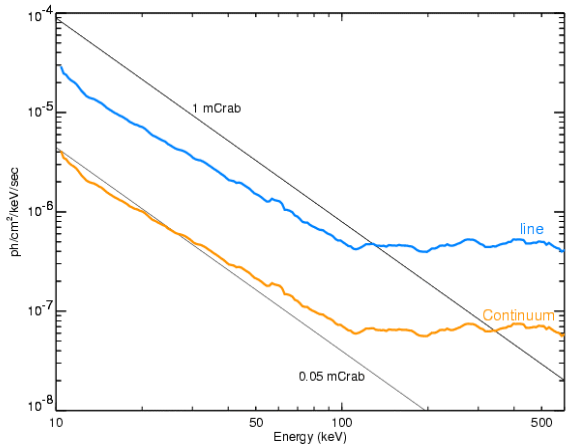

Figure 8. EXIST survey sensitivities (5σ) for ~0.5-1 year exposure and 5mm thick detector.

The survey exposure (Figure 7) is uniform to within ~20-30% (full sky) over a day, with the two maxima corresponding to the orbital poles. These slowly move as the orbit precesses with a ~50d period, but allow a deeper survey with continuous coverage approaching ~100% for a given source for the interval it can be observed near the orbital pole, less reduction due to detector shutdowns during the SAA high background portions of some orbits which for a given source might reduce exposure totals by ~15%. The survey sensitivity shown in Figure 8 assumes a 5mm thick CZT array. The degradation in sensitivity (flattening) above ~100-200 keV would be further improved by a thicker (1cm) detector, although higher operating bias voltages are needed. This option is under study.

Overall parameters for the EXIST mission concept, as currently studied, are given in Table 2.

**Table 2:** EXIST Mission Parameters

| | |
|---|---|
| Energy range | 10-600 keV |
| FOV (instantaneous) | 180° x 75° vs. 5str (fully vs. partially coded) |
| Survey coverage | Full sky each ~95min orbit |
| Angular resolution | 2 - 5' (10-50" source locations) |
| Energy; Temporal resolution | 1-3 keV (FWHM); 2 μsec |
| Sensitivity (5σ) | 10-200 keV: ~0.05 mCrab (≤1y); ~2mCrab (ea. orbit) |
| | 200-600 keV: ~0.5 mCrab (≤1y); ~20mCrab (ea. orbit) |
| Telescopes | Coded aperture: quasi-hemispherical URA mask |
| | 7mm thk. tungsten; 2.5mm pixels radially aligned |
| Detectors | 8m$^2$ CZT (20mm xtals x 5-10mm thk.); 1.25mm pixels |
| Scanning (zenith); pointing | ~1° stability; 5" aspect knowledge |
| Mass; Power; Telemetry | 8500kg; 1500W; 1.5Mbs |
| Launch; Cost (including Ops.) | Delta IV; $380M (including 20% contingency) |

The primary EXIST mission science is the Survey – the highest sensitivity and resolution (both spectral and temporal) imaging coverage of the full sky each ~95min yet obtained at any energy. As such, the mission planning and operations are very simple: continual zenith pointing and mission monitoring, calibration and maintenance. A much more significant effort will be the (near)-real time processing of data and updating of all sky maps, light curves and spectra. The very large data volume (~3 x 10$^4$ events/sec) and processing requirements (each event is in effect back-projected into the collimator-weighted and aspect-corrected ~3 x 10$^5$ sky pixels it "sees" through the coded aperture mask) means that significant software (and possibly hardware) development is needed before launch as well as maintenance throughout the mission.

EXIST will provide the full-sky high energy images to complement many other ongoing missions in space and on the ground. Examples include (see Grindlay et al[13] for discussion) Constellation-X, for which the survey can guide both SXT and, particularly, HXT observations; GLAST, for which both GRB and blazar outburst positions and spectra can be provided; and LSST, with which novae, black hole transients and GRBs can be followed up. To achieve maximum survey depth and temporal coverage, as well as to complement other missions, EXIST is planned as a 5 year mission.

Both its complementarity to other missions, as well as high sensitivity and broad band pass, mean that EXIST will be used as a pointed Observatory for targetted observations (of many sources simultaneously) after the first year of optimally-oriented (Figure 1) survey observations. The principal gain is increased exposure time per source: for sources on the orbital equator, coverage of a given source can be increased from duty cycle ~75/360 to ~180/360 by a 90° rotation of the spacecraft so that the "fan beam" is along, rather than perpendicular to, the orbital ram direction. Sources near the orbital poles are, of course, maximally observed for the normal scanning orientation, and (most) sources at intermediate orbital latitudes can be observed by a slow S/C roll about the Z axis while maintaining inertial pointing with the Upper telescope along the Observatory axis (cf. Figure 2). An option being considered for further increasing sensitivity for such pointed observations is to provide a smaller FoV low energy (10 - ~50 keV) passive collimator for the central ST and surrounding 4 half (above, below; and left, right) and 4 quarter (corner) STs of the Upper Telescope (only) in order to significantly reduce the cosmic diffuse flux which is the dominant source of background. A 10° x 10° central collimator in the central ST and a 10° x 20° collimator, to allow for ST offset angles, in each of the 8 partial STs (i.e. DMs) which surround the central ST (see Figure 3) would reduce backgrounds by a factor of ~15 and thus increase sensitivity for on-axis pointings by a factor of ~4. By providing this low energy collimation in the partial STs (i.e. DMs) only, the loss of low energy sensitivity due to loss of exposure time in the central strip of the Upper Telescope for the scanning survey can be minimized (to a factor of ≤1.5) since the remaining non-collimated DMs still partly provide the original coverage. Detailed simulations are needed, as well as studies of possibly increasing the imaging resolution for this central on-axis "pointed" telescope by using a 2-cycle coded aperture mask[14] in the central part of the Upper Telescope, over this part of the detector array, which could then have 2.5' resolution (1.25mm mask pixels) at low energies while maintaining the 5' resolution at higher energies (≥70-100 keV).

With the estimated sensitivities, and corresponding numbers and types of sources expected[13] (e.g. ~1-4 AGN per degree$^2$), the $\delta\theta$ = 5' (FWHM) resolution is below the canonical confusion limit of one source per 40 beams. The resolution is needed, however, so the centroiding and absolute location capability, $\delta\varphi \sim \delta\theta/(n-3)$ for a source detected with n-sigma significance, is small enough to permit unambiguous source identifications: e.g., a GRB detected with n = 30$\sigma$ significance and thus with flux ~1.5X the BATSE threshold[7], will be located with ~10" (FWHM) centroiding error (and so consistent with the 5" telescope aspect) and thus able (in most cases) to be identified with a particular galaxy from the GRB itself.

Observatory operations would likely be conducted in the second year and beyond of the mission, with Guest Observers proposing for pointings on central targets or other targets within the multi-telescope FoV During such pointed observations, the wide field coverage of the full telescope array enables the Survey to continue. By executing a slow roll about the Z axis to keep the Y axis approximately parallel to the Earth limb when a source on-axis in the Upper telescope is rising or setting, the Left and Right telescopes avoid Earth occultation (and increased backgrounds). All survey data from throughout the full mission would be made public immediately.

## 5. TECHNOLOGY DEVELOPMENT PROGRAM

The mission concept described here for the EXIST freeflyer has been developed in a mission design study conducted at the NASA Goddard Integrated Mission Design Center (IMDC) with significant inputs from research programs at the authors' institutions as well as followup mission design studies at GSFC. Although many more details of the mission have been studied than can be reported here, and the parameters summarized in Table 2 appear to be achievable, a technology development program is needed. The major challenges (see also discussion in §3) to be met over the next ~3-5 years to enable a final mission design are summarized briefly.

**5.1 Detector development**

The largest CZT imaging array yet designed is the BAT detector (0.5m$^2$) now being finished for the Swift mission[15]. This is being constructed with discrete 4 x 4 x 2 mm crystals (planar detectors, without pixel contacts), each connected to the input of an ASIC in groups of 128, for a total of 32768 individually mounted detectors and 256 ASICs. EXIST would incorporate "only" ~20000 CZT crystals, each 2 x 2 x 0.5-1 cm and mounted in groups of 4 on a single DCA (Figure 4) with a single ASIC, but then requires a total of ~5000 ASICs. Thus, the technical challenges are to develop high uniformity and yield, and thus relatively low cost, CZT crystals on the 2cm scale with 1.25mm-pixel contacts as well as the CZT-ASIC mounting and contacting methods. Detector-ASIC prototypes[11] that can be tiled $^{must}$ be extended to smaller pixel size, multi-pixel readout (peak plus neighbors and/or second largest pulse height) and depth sensing capability for Compton event and off-axis imaging reconstruction as noted above. Cathode readout schemes for tileable imaging pixel detectors[12] must be further developed and demonstrated for thick detectors (0.5-1cm) in realistic background (i.e. balloon flight) conditions. Finally, the relatively high bias voltages (~1500V) needed to operate thick CZT detectors efficiently present detector packaging challenges. Development of the very large area CZT imagers needed for EXIST would have important spinoff for both medical imaging (already driving CZT development) and surveillance technologies.

**5.2 ASIC and readout development**

A particularly critical path for EXIST is the development and demonstration of very low power and low noise ASICs for the detector readout. With 8m$^2$ of imaging detector with 1.25mm pixels, there are ~5 x 10$^6$ channels to readout and monitor for gain and offset drifts as well as control for trigger levels. To achieve the total mission power budget of ≤150W per Telescope (each with 1/3 the total detector area), a power of < 50-100mW per channel is needed. The ASIC must be flip-chip mounted onto a coupling board to transfer connections from the CZT pad pitch to the smaller ASIC pixels (see Figure 4) with minimal lead length, capacitance and thus noise. Detector calibration and performance monitoring systems (e.g. low power test pulsers to measure ASIC gains and offsets and tagged external calibration sources) must be developed. Progress in developing prototype low-power/noise ASICs was presented by Harrison et al[16].

**5.3 Coded aperture design and collimation**

Wide-field imaging at relatively high spatial resolution (5') and over a broad energy range (extending to 600 keV) presents challenges when constructed in the compact design envelope dictated by minimizing mission mass. The design concept for EXIST, with 1.5m focal length telescopes, incorporates hemi-spherical or quasi-hemishperical (approximated by flat tiles) coded apertures with radially aligned aperture mask holes to minimize auto-collimation. This novel imaging approach needs further study and optimization with both simulations and prototype mask-imager fabrication and testing. Design studies are needed for the combined active-passive collimator for both shielding and GRB spectroscopy as well as the possible low energy 1D collimator (incorporated in the cathode board design).

**5.4 Data systems and data processing**

The scanning Survey mission (and partial scanning with off-axis telescopes even during Observatory pointings) requires the data to be recorded in event mode with x,y,z position, energy, and shield status, as outlined in §3.4. On board processing of and data formatting will require high speed data bus architectures to be developed, and real-time (~10sec) GRB positions, at an expected rate of ~1-3 per day[7], require high speed on-board computation. Orbit-based processing of transients would require continuous use of the same on board computation resources and must be sized for power and data bus constraints. Full processing of data on the ground presents other interesting challenges, such as dynamically creating all-sky catalogs and source databases for ready access to source spectra and variability, given that coded aperture imaging maps each event onto a vast number of candidate sources as well as the true source.

## 6. CONCLUSIONS

EXIST would conduct highest priority science and a survey unique in Astronomy: full sky imaging on the shortest timescales (~95min). The mission is poised for more detailed design studies and a technology development program to fulfill the major science objectives of a black hole survey and finder mission, as recommended specifically by the Decadal Review and in the context of the "Beyond Einstein" program by the NASA SEU/Roadmap report for which EXIST would compete for the recommended "Black Hole Finder" probe.

## 7. ACKNOWLEDGEMENTS

We thank the EXIST Science Working Group (see http://exist.gsfc.nasa.gov) for input on the mission science and goals, S. Barthelmy, A. Parsons, C. Hailey and G. Fishman for technical assistance in the ISAL and IMDC mission studies at GSFC, R. Carter and S. DePaolo for management of the mission study at GSFC, T. Narita, J. Jenkins and W. Cook for detector and readout design discussions, and T. Hunter for telescope mechanical design assistance. This work was supported in part by NASA SR&T grant NAG5-5729 and support from NASA HQ for the design studies at GSFC is gratefully acknowledged.